\documentclass[notitlepage,nofootinbib,preprintnumbers,amssymb,superscriptaddress]{revtex4-1}
\usepackage{amsfonts,amssymb,amsmath,graphicx,color,bm}
\definecolor{ultramarine}{rgb}{0.07, 0.04, 0.56}
\definecolor{cadmiumgreen}{rgb}{0.0, 0.42, 0.24}
\definecolor{indigo(dye)}{rgb}{0.0, 0.25, 0.42}
\usepackage[linktocpage=true]{hyperref}
\hypersetup{
colorlinks=true,
citecolor=ultramarine,
linkcolor=cadmiumgreen,
urlcolor=indigo(dye),
}

\newcommand{\f}[2]{\frac{#1}{#2}}  
\newcommand{\mk}[1]{\left( #1 \right)}  
\newcommand{\kk}[1]{\left[ #1 \right]}  
\newcommand{\ck}[1]{\left\{ #1 \right\}}  
\newcommand{\be}{\begin{equation}}  
\newcommand{\ee}{\end{equation}}
\newcommand{\Mpl}{M_{\rm Pl}}

\begin{document}%%%%%%%%%%%%%%%%%%%%%%%%%%%%%%%%%%%%%%%%%  

\title{  
Constant-roll inflation: confrontation with recent observational data
}

\author{Hayato Motohashi}
%\email[E-mail:]{motohashi"at"ific.uv.es}
\affiliation{Instituto de F\'{i}sica Corpuscular (IFIC), Universidad de Valencia-CSIC, E-46980, Valencia, Spain}
\affiliation{Research Center for the Early Universe (RESCEU),
Graduate School of Science, The University of Tokyo, Tokyo 113-0033, Japan}
\author{Alexei A.\ Starobinsky}
%\email[E-mail:]{alstar"at"landau.ac.ru} 
\affiliation{Research Center for the Early Universe (RESCEU),
Graduate School of Science, The University of Tokyo, Tokyo 113-0033, Japan}
\affiliation{L. D. Landau Institute for Theoretical Physics RAS, Moscow 119334, Russia}

\begin{abstract}%%%%%%%%%%%%%%%%%%%%%%%%%%%%%%%%%%%%%%%%%  
The previously proposed class of phenomenological inflationary models
in which the assumption of inflaton slow-roll is replaced by the
more general, constant-roll condition is compared with the most 
recent cosmological observational data, mainly the Planck ones.
Models in this two-parametric class which remain viable appear
to be close to the slow-roll ones, and their inflaton potentials
are close to (but still different from) that of the natural inflation
model. Permitted regions for the two model parameters are
presented.
\end{abstract}

\maketitle  

%%%%%%%%%%%%%%%%%%%%%%%%%%%%%%%%%%%%%%%%%  

\section{Introduction}%%%%%%%%%%%%%%%%%%%%%%%%%%%%%%%%%%%%%%%%%
\label{sec-int}

The dramatic increase in accuracy of cosmological observational data occurred during the last few years, among which the most important for the inflationary scenario of the early Universe are the value of the slope $n_s$ of the Fourier power spectrum of primordial scalar (adiabatic matter density) perturbations $P_{\zeta}(k)$ and the upper value on the tensor-to-scalar ratio $r(k)$, see \cite{Array:2015xqh} for the most recent data, gives the possibility to go beyond some of the main assumptions on which all existing viable inflationary models are based and then to use the data to see to what extent they do permit it. In particular, in this paper we want to abandon the standard slow-roll approximation for inflaton motion during inflation which guarantees both smallness of scalar and tensor perturbations generated during inflation and ``graceful exit'' from it.\footnote{Note that the slow-roll approximation is not specific for inflation only. In fact, it was first used before the development of the inflationary scenario, in the context of its main rival -- a bouncing universe, in \cite{Star78} for the $V=m^2\phi^2/2$ scalar field potential in a closed Friedmann-Lema\^itre-Robertson-Walker (FLRW) universe where it first resulted in "slow climb" of the scalar field during contraction and then slow roll after a bounce.} Of course, this can be done in many ways. We shall study the specific, but natural and elegant generalization of inflation driven by a slow-rolling scalar field in General Relativity (GR) -- constant-roll inflation.

This is a two-parametric class of phenomenological inflationary models in GR in which the condition of a constant rate of the inflaton motion, see Eq.~(\ref{cr-con}) below, is satisfied {\em exactly}. Then, neglecting spatial curvature and other types of matter apart from the inflaton $\phi$ itself and using the Hamilton-Jacobi-like equation for $H(\phi)$~\cite{Muslimov:1990be,Salopek:1990jq} where $H=d\ln a(t)/dt$ is the Hubble function, $a(t)$ is the scale factor of a FRLW universe, it is straightforward to derive the scalar field potential $V(\phi)$ needed for the constant-roll condition to be satisfied.

The attempt of such a generalization first proposed in \cite{Martin:2012pe}, and the inflaton potential was constructed so that it satisfied the constant-roll condition approximately. Later in \cite{Motohashi:2014ppa}, the generic potential satisfying the constant-roll condition exactly was found which has a rather simple and elegant form, see Eq.~(\ref{crpot}) below for one of its three possible types which is most interesting from the observational point of view. Thus, it is possible to expect that it may appear in some more fundamental theory. The potential is novel by itself, though for some of its types and values of its parameters, it can be close to that in models studied previously, in particular, power-law inflation and natural inflation. In addition, it was found in~\cite{Motohashi:2014ppa} in which cases the constant-roll solution represents an attractor for other inflationary solutions and when curvature perturbation are approximately constant on super-Hubble scales (that is not guaranteed generically for non-slow evolution of background due to mixing with the other, `decaying' mode of them).

The aim of this paper is to move constant-roll inflation closer to reality. Thus, in Sec.~\ref{sec-back} we numerically calculate the power spectrum of scalar (matter density) perturbations generated during it by using the exact solution for background, compare it with the most recent observational data and find a viable region for the two model parameters. Final conclusions are presented in Sec.~\ref{sec-conc}.

\section{Constant-roll inflation}%%%%%%%%%%%%%%%%%%%%%%%%%%%%%%%%%%%%%%%%%
\label{sec-back}

The constant-roll condition has the form~\cite{Martin:2012pe,Motohashi:2014ppa}\footnote{Note that $\beta$ here is related to the parameter $\alpha$ used in \cite{Motohashi:2014ppa} through $\beta=-(3+\alpha)$.}
\be \ddot \phi=\beta H\dot \phi.  \label{cr-con} \ee
So, the slow-roll approximation corresponds to $|\beta|\ll 1$. Now we want to determine the range of $\beta$ permitted by observations without assuming it to be small in advance. We consider only one of the three possible types of constant-roll potentials found in~\cite{Motohashi:2014ppa},  
\be \label{crpot} V(\phi)= 3M^2\Mpl^2\kk{1-\f{3+\beta}{6}\ck{1-\cos \mk{\sqrt{2\beta } \f{\phi}{\Mpl} } } } , \ee 
since it has been already shown there that two other types (with purely exponential and hyperbolic potentials) are already excluded by the data.
This potential differs from that assumed in the case of natural inflation~\cite{Freese:1990rb} by an additional negative cosmological constant. As discussed below, this requires to cut it at some critical value $\phi=\phi_0$ to get an exit from inflation (similar to the case of power-law inflation). The corresponding exact solution for $a(t), H(t)$ and $\phi(t)$ is
\begin{align}
\phi&=  2\sqrt{\f{2}{\beta }}\Mpl {\rm arctan} (e^{\beta Mt}), \\
H&= -M\tanh \mk{\beta Mt} = M\cos\mk{ \sqrt{\f{\beta }{2}} \f{\phi}{\Mpl} } , \\
a&\propto \cosh^{-1/\beta } \mk{\beta Mt} = \sin^{1/\beta } \mk{ \sqrt{\f{\beta }{2}} \f{\phi}{\Mpl} }  .
\end{align}
We consider $\beta \gtrsim 0$ that leads to a red-tilted spectrum since in the opposite case, as was shown in ~\cite{Motohashi:2014ppa}, the constant-roll solution is not an attractor and curvature perturbations on super-Hubble scales experience an anomalous growth that makes the model not feasible observationally.

The critical value $\phi_c$ where $V=0$ is given by
\be \phi_c = \f{\Mpl}{\sqrt{2\beta }} \arccos \mk{1-\f{6}{3+\beta}}, \ee
which is $\phi_c/\Mpl=14.9$ and $21.4$ for $\beta=0.02$ and $0.01$, respectively.
We need to cut the potential before $V<0$, thus the cutoff field position $\phi_0$ should satisfy $\phi_0<\phi_c$ for a given value of $\beta$.
For $\phi_0 \ll \phi_c$ the model looks like the quadratic hilltop inflation with cutoff, whereas for $\phi_0 \lesssim \phi_c$ the model looks like the natural inflation with the additional negative cosmological constant $\Lambda=M^2(3+\alpha)$ as was pointed above.

By using the exact solution, we can evaluate a field position $\phi_i$, which we set is $55$ e-folds back from $\phi_c$. 
We obtain $\phi_i/\Mpl=3.38$ and $8.68$ for $\beta=0.02$ and $0.01$, respectively.
We can choose the cutoff $\phi_0$ by either $\phi_0 \ll \phi_c$ or $\phi_0 \lesssim \phi_c$, 
but in both cases generation of perturbations of our interest takes place within $0<\phi<\phi_i$.
Therefore, we can investigate both cases at the same time, and the region $0<\phi<\phi_i$ includes all possible choices of $\phi_0$.

The potential slow-roll parameters are given by
\begin{align}
\epsilon &\equiv \f{1}{2} \mk{\f{V'}{V}}^2  =  \f{ \beta (3+\beta)^2 \sin^2(\sqrt{2\beta } \phi/\Mpl) }{ [ -6-\alpha+\alpha \cos (\sqrt{2\beta } \phi/\Mpl) ]^2}, \\
\eta&\equiv \f{V''}{V} = \f{ 2\beta (3+\beta) \cos (\sqrt{2\beta } \phi/\Mpl) }{ -3+\beta-(3+\beta) \cos (\sqrt{2\beta } \phi/\Mpl) },\\
\xi&\equiv \f{V'V''}{V^2} = - \f{ 4 \beta^2 (3+\beta)^2 \sin^2(\sqrt{2\beta } \phi/\Mpl) }{[ -3+\beta-(3+\beta) \cos (\sqrt{2\beta } \phi/\Mpl) ]^2 }  ,
\end{align}
which are at most $O(10^{-2})$ for $0.005<\beta<0.025$ and $0<\phi<\phi_i$.
Therefore, for this parameter region the slow-roll approximation applies, and 
the spectral parameters are well approximated by the potential slow-roll parameters as 
\begin{align}
n_s-1 &= -6\epsilon+2\eta, \\
r&= 16\epsilon,\\
\f{dn_s}{d\ln k}&= 16\epsilon \eta-24\epsilon^2-2\xi,
\end{align}
where $n_s$ is the spectral index of the scalar power spectrum and $r$ is the ratio between tensor and scalar power spectrum.

\begin{figure}[t]
	\centering
	\includegraphics[width=100mm]{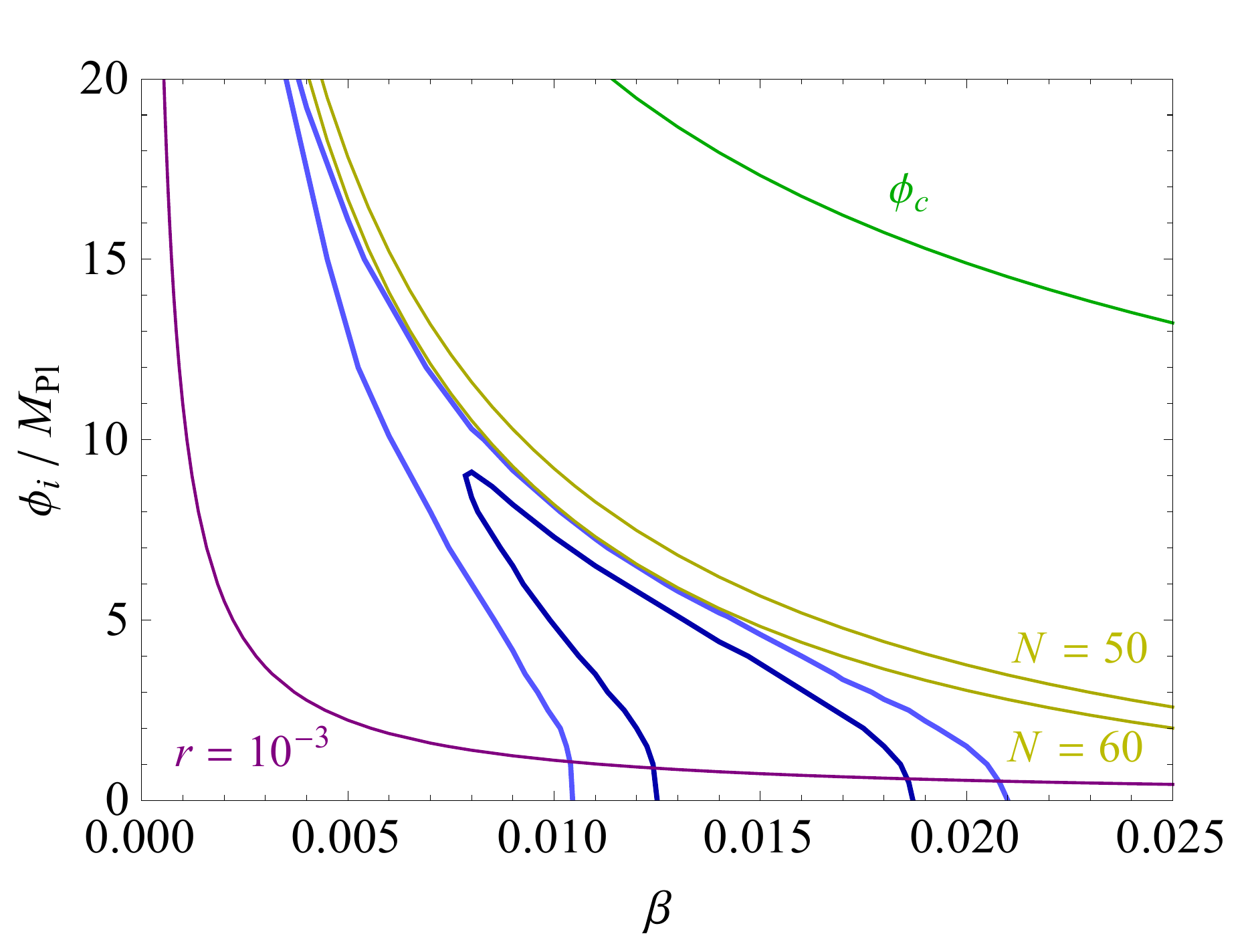}
	\caption{Observational constraint on model parameter space $(\beta,\phi_i/\Mpl)$ of constant-roll inflation.  $68\%$ and $95\%$ confidence regions (blue), $\phi_c$ (green), field position for $50$ e-folds and $60$ e-folds back from $\phi_c$ (yellow), and $r=10^{-3}$ (purple).}
	\label{fig:contour}
\end{figure}

We exploit the consistency relation for $n_s$ and $r$ with the observational constraint from the joint analysis of Planck and BICEP2/Keck Array, namely, Fig.~7 in \cite{Array:2015xqh}, and obtain the constraint in Fig.~\ref{fig:contour} for model parameters $(\beta,\phi_i/\Mpl)$.  
Blue curves represent $68\%$ and $95\%$ confidence regions, which are within $0<\phi<\phi_i$.
Actually, these regions are more than $60$ e-folds back from $\phi_c$.  
We also plotted a purple curve for $r=10^{-3}\approx (1-n_s)^2$ for which the hope to reach it in a not so remote future is realistic, see e.g.~\cite{Andre:2013nfa}.

%==================== Figure ====================
%==================== Figure ====================

\section{Conclusion}%%%%%%%%%%%%%%%%%%%%%%%%%%%%%%%%%%%%%%%%%
\label{sec-conc}

We have investigated most recent observational constraints on parameters of generic constant-roll inflationary models, in which the standard assumption of inflaton slow-roll is replaced by the more general, constant-roll condition. The inflaton potential producing such exact background behaviour is given, in the case permitted by observations, by a sinusoidal curve minus a small constant which is close to but still different from that in the natural inflation model. The corresponding exact analytic solution for the background space-time metric and the inflaton field appears to be an attractor for inflationary dynamics. Using the most recent cosmological observational constraints on the spectral index $n_s$ and tensor-to-scalar ratio $r$ obtained in \cite{Array:2015xqh}, we have found a viable region for the two parameters of the model presented in Fig.~\ref{fig:contour}. It appears that though we did not assume the parameter $\beta$ to be small in advance, the data tell us that it should lie in the range $(0.01, 0.02)$ approximately -- which is small indeed. However, the exact background constant-roll solution is still useful since it provides us with the possibility to calculate small corrections to the slow-roll result in any order of the perturbation theory in slow-roll parameters and to check the convergence of the latter one. As for expected values of the tensor-to-scalar ratio $r$, the allowed region for them ranges from values as large as those permitted by the present upper limits ($r\lesssim 0.07$) up to very small ones $r\ll (1-n_s)^2 \approx 10^{-3}$ in the small-field case $\phi_i\ll \Mpl $. However, for a sufficiently large area of allowed model parameters corresponding to large-field inflation with $\phi_i > \Mpl$, $r$ exceeds $10^{-3}$ which will be accessible in near-future observations.

\begin{acknowledgments}%%%%%%%%%%%%%%%%%%%%%%%%%%%%%%%%%%%%%%%%%
We thank the Research Center for the Early Universe, where part of this work was completed.  
H.M.\ is supported in part by MINECO Grant SEV-2014-0398,
PROMETEO II/2014/050,
Spanish Grant FPA2014-57816-P of the MINECO, and
European Union’s Horizon 2020 research and innovation programme under the Marie Sk\l{}odowska-Curie grant agreements No.~690575 and 674896.
A.S.\ acknowledges RESCEU for hospitality as a visiting professor. He was also partially supported by the grant RFBR 17-02-01008 and by the Scientific Programme P-7 (sub-programme 7B) of the Presidium of the Russian Academy of Sciences. 
\end{acknowledgments}%%%%%%%%%%%%%%%%%%%%%%%%%%%%%%%%%%%%%%%%%

\bibliography{ref-conroll2}

\end{document}